\documentclass{PoS}

\usepackage{graphicx}
\usepackage{amsmath}
\usepackage{subfigure}
\usepackage{wrapfig}
\usepackage{epsfig}
\def\lsim{\raise0.3ex\hbox{$<$\kern-0.75em\raise-1.1ex\hbox{$\sim$}}}
\def\gsim{\raise0.3ex\hbox{$>$\kern-0.75em\raise-1.1ex\hbox{$\sim$}}}

\newcommand{\be}{\begin{equation}}
\newcommand{\ee}{\end{equation}}
\newcommand{\bea}{\begin{eqnarray}}
\newcommand{\eea}{\end{eqnarray}}

\newcommand{\Tr}{\mathop{\textrm{Tr}}}
\newcommand{\exv}[1]{\left\langle{#1}\right\rangle}
\newcommand{\x}{{\bf x}}
\renewcommand{\r}{{\bf r}}

\title{QCD Thermodynamics}
\ShortTitle{QCD Thermodynamics}

\author{Zoltan Fodor\\
John von Neumann Institute for Computing (NIC), DESY, D-15738, Zeuthen /
FZJ, D-52425, Juelich, Germany\\
 Department of Physics, University of Wuppertal, Gauss Strasse 20, D-42119, Wuppertal, Germany\\
Institute for Theoretical Physics, Eotvos University, Pazmany 1, H-1117 Budapest, Hungary

 Email: \email{fodor@bodri.elte.hu}}

\abstract{
Recent results on QCD thermodynamics are presented. The nature of the 
T$>$0 transition is determined, which turns out to be an analytic 
cross-over. The absolute scale for this transition is calculated. 
The temperature dependent static potential is given. The results were
obtained by using a Symanzik improved gauge and stout-link improved
fermionic action. In order to approach the continuum limit four different
sets of lattice spacings were used with temporal extensions $N_t$=4, 6, 8 and 
10 (they correspond to lattice spacings $a$$\sim$0.3, 0.2, 0.15 and 
0.12~fm). A new technique is presented, which --in contrast to earlier
methods--  enables one to determine the equation of state at 
very large temperatures. 
}

\FullConference{The XXV International Symposium on Lattice Field Theory\\
		 July 30 - August 4 2007\\
		 Regensburg, Germany}

\begin{document}

\section{Introduction}
The QCD transition at non-vanishing temperatures ($T$) 
plays an important role in the physics of
the early Universe and of heavy ion collisions (most recently at RHIC at
BNL; LHC at CERN and FAIR at
GSI will be the next generation of machines). 
The main goal of the present summary is to present results of the
Budapest-Wuppertal group on the QCD transition 
at vanishing chemical potential ($\mu$=0), which is of direct
relevance for the early universe ($\mu$ is negligible there) and for
present and future heavy ion collisions (at RHIC $\mu$$\lsim$40~MeV, which is
far less than the
typical hadronic scale). Since these results were obtained at four different
sets of lattice spacings and a careful continuum extrapolation was performed,
we consider them as full results. 
In addition, a new technique will be introduced, 
which closes the gap between lattice QCD and the perturbative approach
for bulk thermodynamical quantities.

\FIGURE[t]{
\includegraphics[width=5.5cm,angle=0]{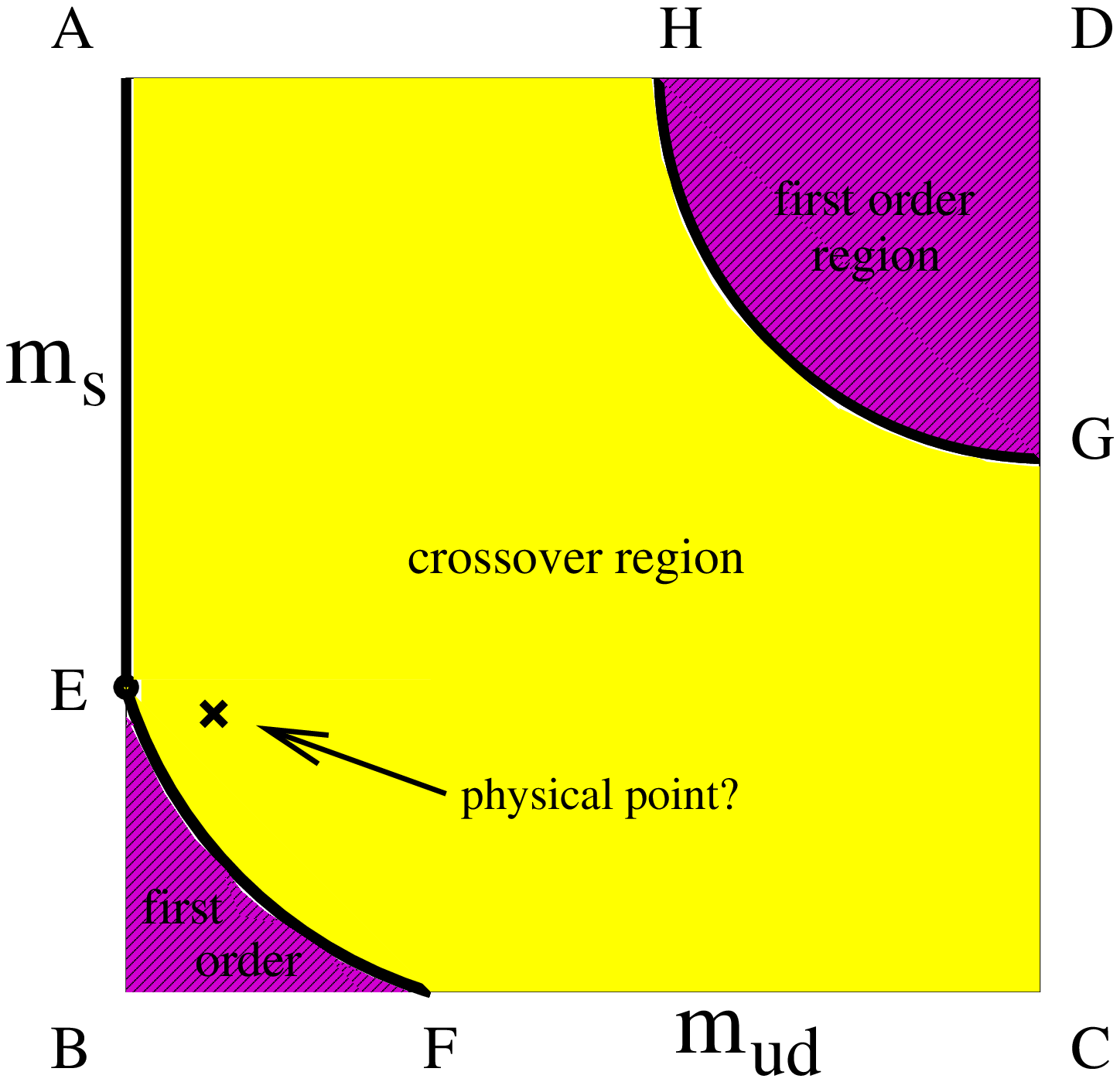}
\caption{\label{phase-ud_s}
The phase diagram of QCD on the hypothetical light quark mass versus
strange quark mass plane. Thick lines correspond to second order
phase transitions, the purple regions represent first order phase 
transitions and the yellow region represents an analytic cross-over.
}
}

The standard picture for the QCD phase diagram on the light quark mass 
($m_{ud}$) versus strange quark mass ($m_s$) plane is shown by Figure 
\ref{phase-ud_s}. It 
contains two regions at small and at large quark masses, for which the $T>0$
QCD transition is of first order. Between them one finds
a cross-over region, for which the $T>0$ QCD transition is an analytic one.   
The first order transition regions and the cross-over region are separated
by lines, which correspond to second order phase transitions.

When we analyze the nature and/or the absolute scale of the $T>0$ QCD 
transition for the physically relevant case two ingredients
are quite important. 

First of all, one should use physical quark masses. 
As Figure \ref{phase-ud_s} 
shows the nature of the transition depends on the quark mass,
thus for small or large quark masses it is a first order phase transition, 
whereas for intermediate quark masses it is an analytic cross over. Though
in the chirally broken phase chiral perturbation theory provides a 
controlled technique to gain information for the quark mass dependence, it
can not be applied for the $T>0$ QCD transition (which deals with the
restoration of the chiral symmetry). In principle, the behaviour of 
different quantities in the critical region 
(in the vicinity of the second order phase transition line) might give
some guidance. However, a priori it is not known how large this region is.
Thus, the only consistent way to eliminate uncertainties related to 
non-physical quark masses is to use physical quark masses (which is, of 
course, quite CPU demanding).

\begin{figure}\begin{center}
{\includegraphics[width=6.4cm,height=5.9cm]{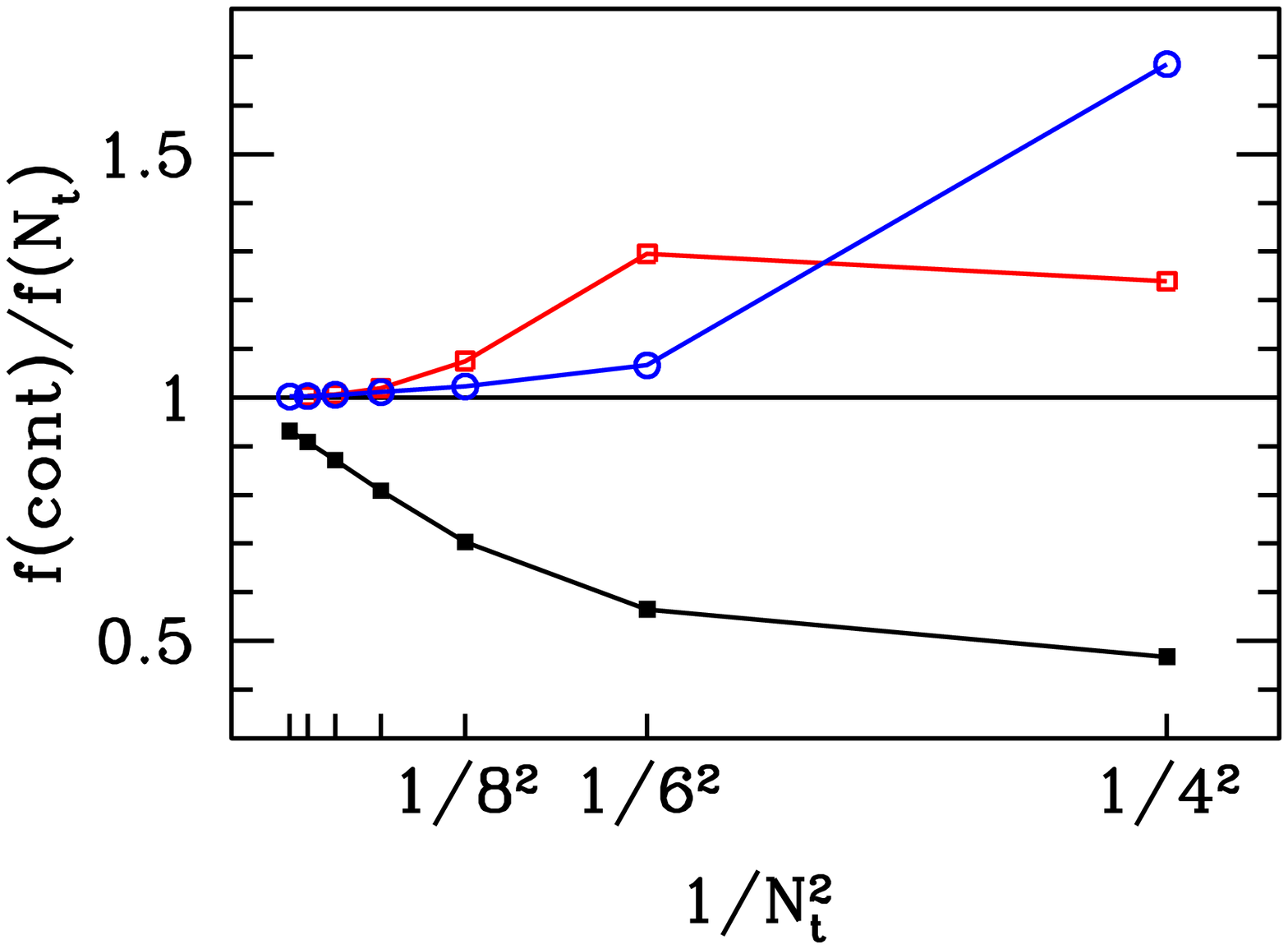}\hspace{.3cm}\includegraphics[width=6.3cm,height=5.9cm]{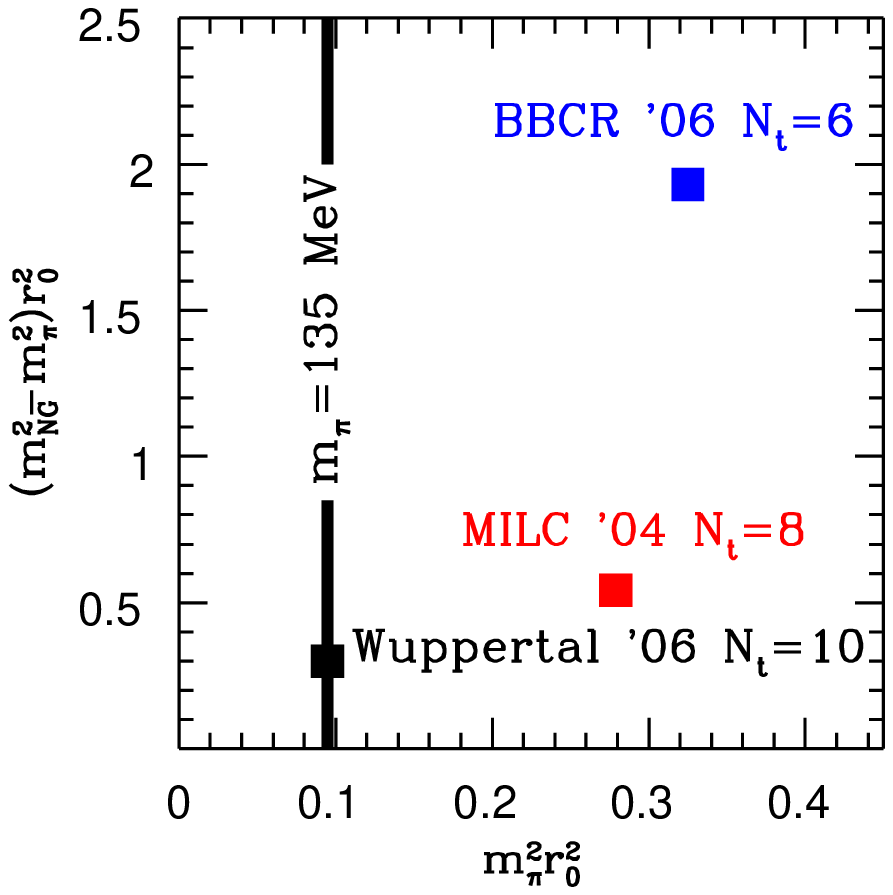}}
\end{center}\caption{\label{advantages}
The ratio $f(cont)$/$f(N_t)$ as a function of $1/N_t^2$ (left panel). 
$f(cont)$ is the continuum extrapolated free energy of the staggered 
fermionic gas 
in the non-interactive, infinitely high temperature limit. $f(N_t)$ is 
the value obtained on a lattice with $N_t$ temporal extension. The black line
shows our choice (stout improvement, only next-neighbour terms in the action),
whereas the red and blue lines represent the Naik and p4 actions, respectively.
Masses and taste symmetry violation for different approaches in the 
literature (right panel). 
The smallest, physical quark mass and the smallest taste symmetry
violation was reached by our works 
(black dot, \cite{Aoki:2006we,Aoki:2006br}). Somewhat larger taste symmetry
violation and about three times larger quark masses were reached by
the MILC analysis on QCD thermodynamics
(red dot, \cite{Bernard:2004je}). Even larger taste symmetry
violation and about four times the physical quark masses are the
characteristics of the Bielefeld-Brookhaven-Columbia-Riken result on 
$T_c$ (blue dot,
\cite{Cheng:2006qk}).
}
\end{figure}

Secondly, the nature of the $T>0$ QCD transition is known to suffer from
discretization errors 
\cite{Karsch:2003va,Philipsen:2007,Endrodi:2007}. The three
flavour theory with
standard action on $N_t$=4 lattices predicts a critical pseudoscalar mass 
of about 300~MeV. This point separates the first order 
and cross-over regions of Figure 1. If we
took another discretization, with another discretization error, 
e.g. the p4 action and $N_t$=4, the
critical pseudoscalar mass turns out to be around 70~MeV (similar effect is
observed if one used stout smearing improvement and/or $N_t$=6). 
Since the physical
pseudoscalar mass (135~MeV) is just between these two values, the discretization
errors in the first case would lead to a first order transition, whereas 
in the second case to a cross-over. 
The only way to resolve this inconclusive situation is to carry out a 
careful continuum limit analysis. 

Since the nature of the transition influences the absolute
scale ($T_c$) of the transition --its value, mass dependence, uniqueness 
etc.-- the above comments are valid for the determination of $T_c$, too.
 
Therefore, we have to answer the question: what
happens for physical quark masses, in the continuum, at what $T_c$? In order to
get a reliable answer we
used physical quark masses on
$N_t$=4,6,8 and 10 lattices, which correspond to approximately 0.3, 0.2, 
0.15
and 0.12~fm lattice spacings, respectively. 

In the presentation \cite{Karsch:2007} of the results of the  
Bielefeld-Brookhaven-Columbia-Riken Collaboration published
results \cite{Cheng:2006qk} 
from $N_t$=4 and 6 were shown (and some unpublished figures for
$N_t$=8, which were obtained within the HotQCD Collaboration).  
Since the CPU requirements for thermodynamics
increase as  $\approx N_t^{12}$ our $N_t$=10 simulations need about 50 times 
more CPU than
$N_t$=6. Do we have 50 times more resources for QCD thermodynamics 
than our competitors? Of
course not (it is almost the other way around). Instead, reaching $N_t$=10 
is a fine balance. It is partly related to
the choice of our action (which will be discussed in the next section), 
partly to the arrangements of the financial resources.
For instance, as $N_t$ increases, one needs more and more statistics. Thus
the thermalization can be done only once on a relatively expensive,
scalable machine, such as
Blue-Gene/L, whereas a large fraction of the non-vanishing T simulations can
be done on more cost effective devices such as personal computer 
\cite{Egri:2006zm} graphics cards.
A 2 years old model can accommodate $N_t$=6 lattices, on a one-and-a-half
year old model you can put $N_t$=8 lattices and the one year old model
can work with quite large $N_t$=10 lattices. It costs a few hundred dollars
and can provide upto 30--60~Gflops sustained QCD performance. 
They are not easy to code, adding two
numbers needs 3 pages, but recently more efficient programming environments
were introduced. Clearly, this type of hardware provides a very advantageous
price--performance ratio for lattice QCD.

\section{The choice of the action}

\begin{figure}\begin{center}
{\includegraphics[width=15cm,bb=45 545 570 690]{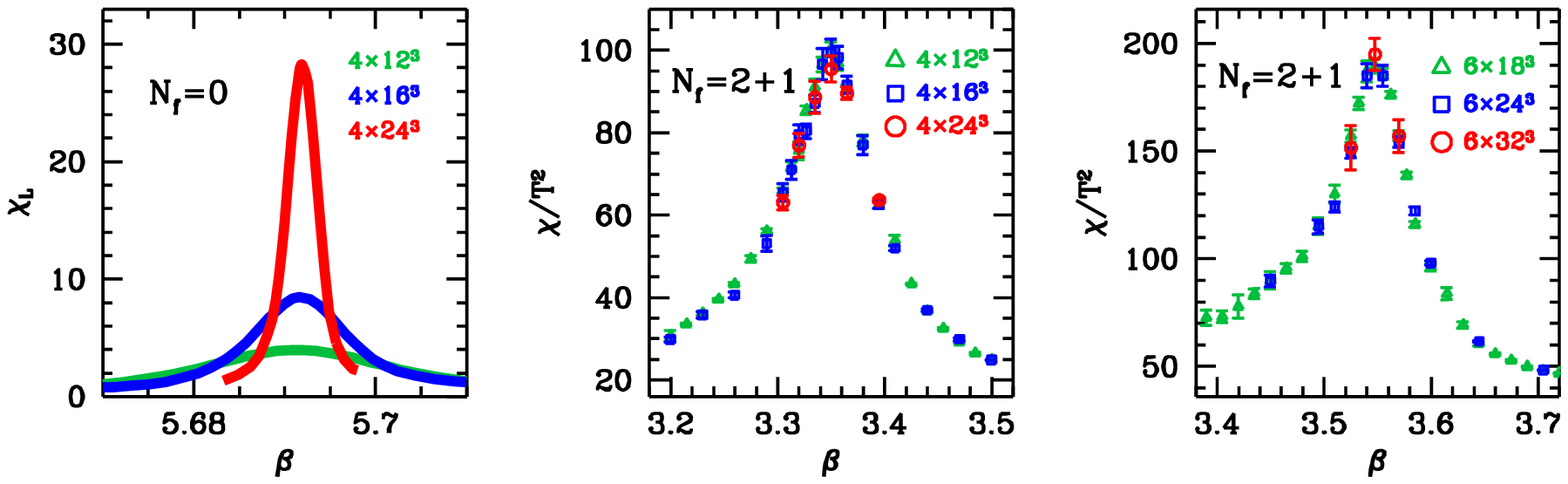}}
\end{center}\caption{\label{susc_446}
The volume dependence of the susceptibility peaks for pure SU(3) gauge 
theory (Polyakov-loop susceptibility, left panel) and for full QCD (chiral
susceptibility on $N_t$=4 and 6 lattices, middle and right panels, 
respectively). 
}
\end{figure}

The first step is to choose an action, which respects all the needs of 
a thermodynamic analysis. T=0 simulations are needed to set the scale and
for renormalization. T>0 simulations are needed to map the behaviour
of the system at non-vanishing temperatures. The action should maintain
a balance between these two needs, leading to approximately the same
uncertainties for both sectors (otherwise a large fraction of the CPU-power
is used just for ``over-killing'' one of the two sectors). We used
Symanzik improved gauge and stout improved 2+1 flavour staggered
fermions \cite{Aoki:2005vt} (due to the stout improvement we have only
next-neighbour terms in the fermionic part of the action). 
The simulations were done along the line of constant physics.
The parameters were tuned with a quite high precision,
thus at all lattice spacings the $m_K/f_k$ and $m_K/m_\pi$ ratios
were set to their experimental values with an accuracy better than 2\%.

\begin{figure}\begin{center}
\includegraphics[width=6.5cm,angle=0,bb=18 180 570 605]{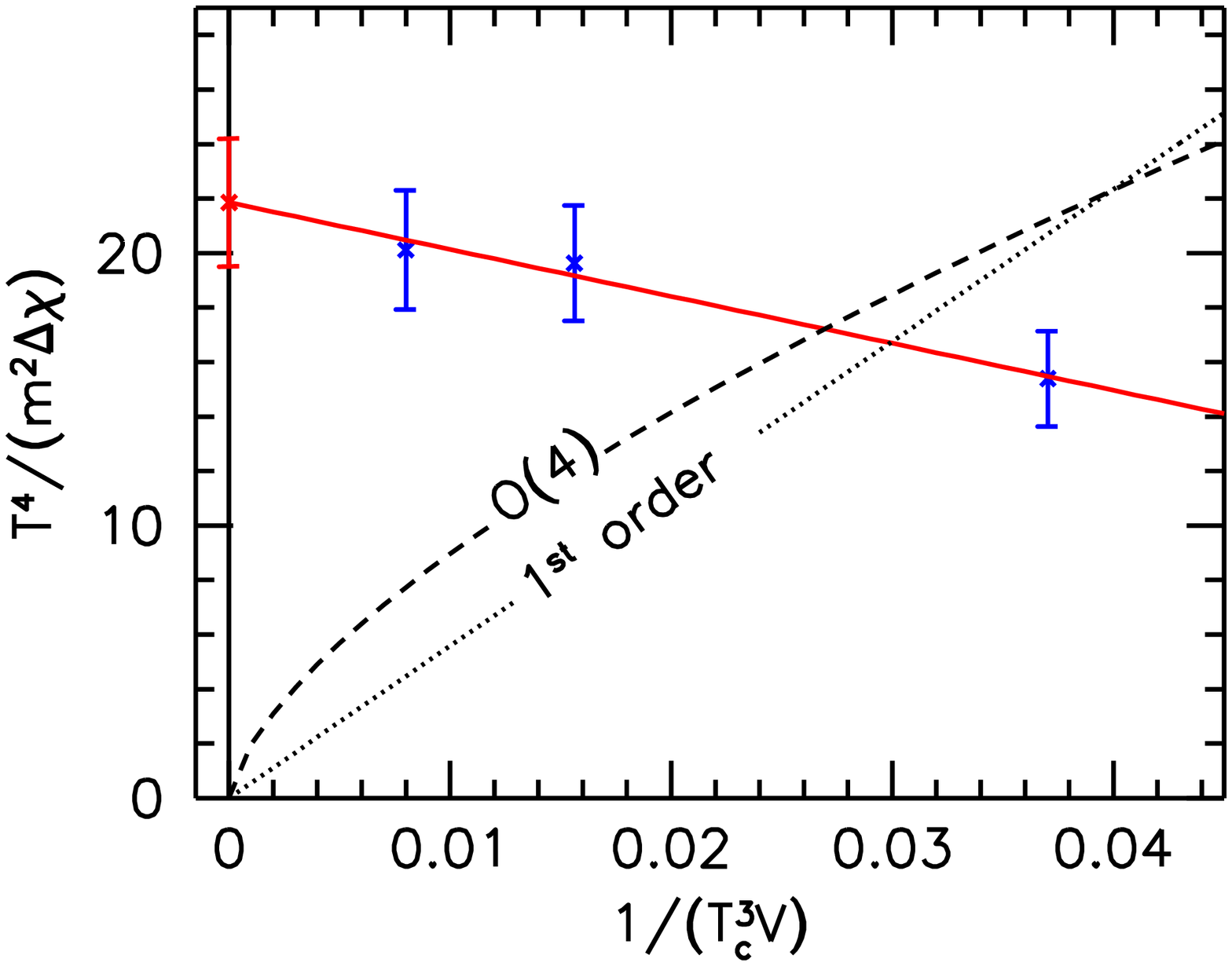}
\end{center}
\caption{\label{cont_scal}
Continuum extrapolated susceptibilities $T^4/(m^2\Delta\chi)$
as a function of 1/$(T_c^3V)$.  For true phase transitions the infinite 
volume extrapolation should be consistent with zero, whereas for an 
analytic crossover the infinite volume extrapolation gives a 
non-vanishing value. The continuum-extrapolated susceptibilities show no 
phase-transition-like volume dependence, though the volume changes by a 
factor of five.
The V$\rightarrow$$\infty$ extrapolated value is 22(2) which is
11$\sigma$ away from zero. For illustration, we fit the expected
asymptotic behaviour for first-order and O(4) (second order)
phase transitions shown by dotted and dashed lines, which results in 
chance probabilities of $10^{-19}$ ($7\times10^{-13}$), respectively. 
}
\end{figure}

The choice of the action has
advantages and disadvantages. As we will see the advantages are probably 
more important than the disadvantages. The left panel of Figure \ref{advantages}
shows the continuum free energy devided by its value at a given $N_t$.
A related plot is usually shown by the Bielefeld-Brookhaven-Columbia-Riken 
collaboration as a function of $N_t$. 
Since in staggered QCD most
lattice corrections scale with $a^2$, which is proportional to 1/$N_t^2$, 
it is instructive to show 
this ratio as a function of 1/$N_t^2$, for our action, for Naik and for p4.
Extrapolations from 4 and 6 always overshoot or undershoot. Clearly, the Naik
and p4 actions reach the continuum value much faster than our choice, 
but the $a^2$ scaling 
appears quite early even for actions with next-neighbour interactions. 
Extrapolations from $N_t$ and $N_t$+2 with our action are 
approximately as good as $N_t$ with the p4 action (which was 
 taylored to be optimal for this quantity, namely for the free 
energy at infinitely large temperatures). 
In practice, it means that our choice with $N_t$=8,10 gives approximately
2\% error for the free energy. In a balanced analysis you do not need more, 
because the corresponding lattice
spacings 0.15 and 0.12~fm are most probably not fine enough to set the scale 
unambiguously with the
same accuracy. (E.g. the asqtad action at $N_t$$\approx$10, which 
corresponds about a=0.12~fm lattice spacing, still has $\approx$10\% scale
difference between $r_1$ \& $f_K$.) 
Since the p4 action is almost 20 times more expensive than
the stout action, it is not worth to pay this price and improve one part of
the calculation, which hinders you to reach a reasonable accuracy in another
part of the calculation. 

This is the balance one should remember in
thermodynamics. Indeed, 
taste symmetry violation should be suppressed for many reasons
(setting the scale at $T=0$, restoring chiral symmetry at $T>0$ etc.) As it was argued
above it is more important to improve on this sector of the calculation than
on the infinitely high temperature behaviour. 
The left panel of Figure \ref{advantages} shows the
splitting between the Goldstone and the first non-Goldstone pions for the
Bielefeld-Brookhaven- Columbia-Riken Collaboration (which is far beyond
their kaon mass), for MILC and for us. Our small splitting is partly related
to the stout improvement and partly to the cost issues, since smaller
lattice spacings could be used, which resulted in smaller splitting. 
In addition, the cost issues allowed us to use
in our finite T simulations physical quark masses instead of much larger
masses.

\section{The nature of the QCD transition}

The next topic to be discussed is the nature of the QCD transition. 
Physical quark masses were used and a continuum extrapolation
was carried out by using four
different lattice spacings. The details of the calculations can be
found in \cite{Aoki:2006we}. 
In order to determine the nature of the transition one should apply finite size
scaling techniques for the chiral susceptibility
$\chi=(T/V)\cdot (\partial^2\log Z/\partial m_{ud}^2)$. 
This quantity shows a pronounced peak
as a function of the temperature. For a first order phase transition, such
as in the pure gauge theory, the peak of the analogous Polyakov
susceptibility gets more and more singular as we increase the volume (V). The
width scales with 1/V the height scales with volume (see left panel of Figure
\ref{susc_446}). 
A second order transition shows a
similar singular behaviour with critical indices. 
For an analytic transition (what we call a
cross-over) the peak width and height saturates to a constant value. 
That is what we observe in full QCD on $N_t$=4 and 6 lattices (middle and
right panels of Figure \ref{susc_446}). We see an order of magnitude difference
between the volumes,
but a volume independent scaling. It is a clear indication for a cross-over. 
These results were obtained with physical quark masses for two sets of 
lattice spacings. Note, however, that for a final conclusion the important
question remains: do we get the same volume independent scaling 
in the continuum; 
or we have the unlucky case what we had in the Introduction for 3 flavour QCD 
(namely the
discretization errors changed the nature of the transition for the physical
pseudoscalar mass case)?

\FIGURE[t]{
\includegraphics[width=6.5cm,bb=0 180 570 620]{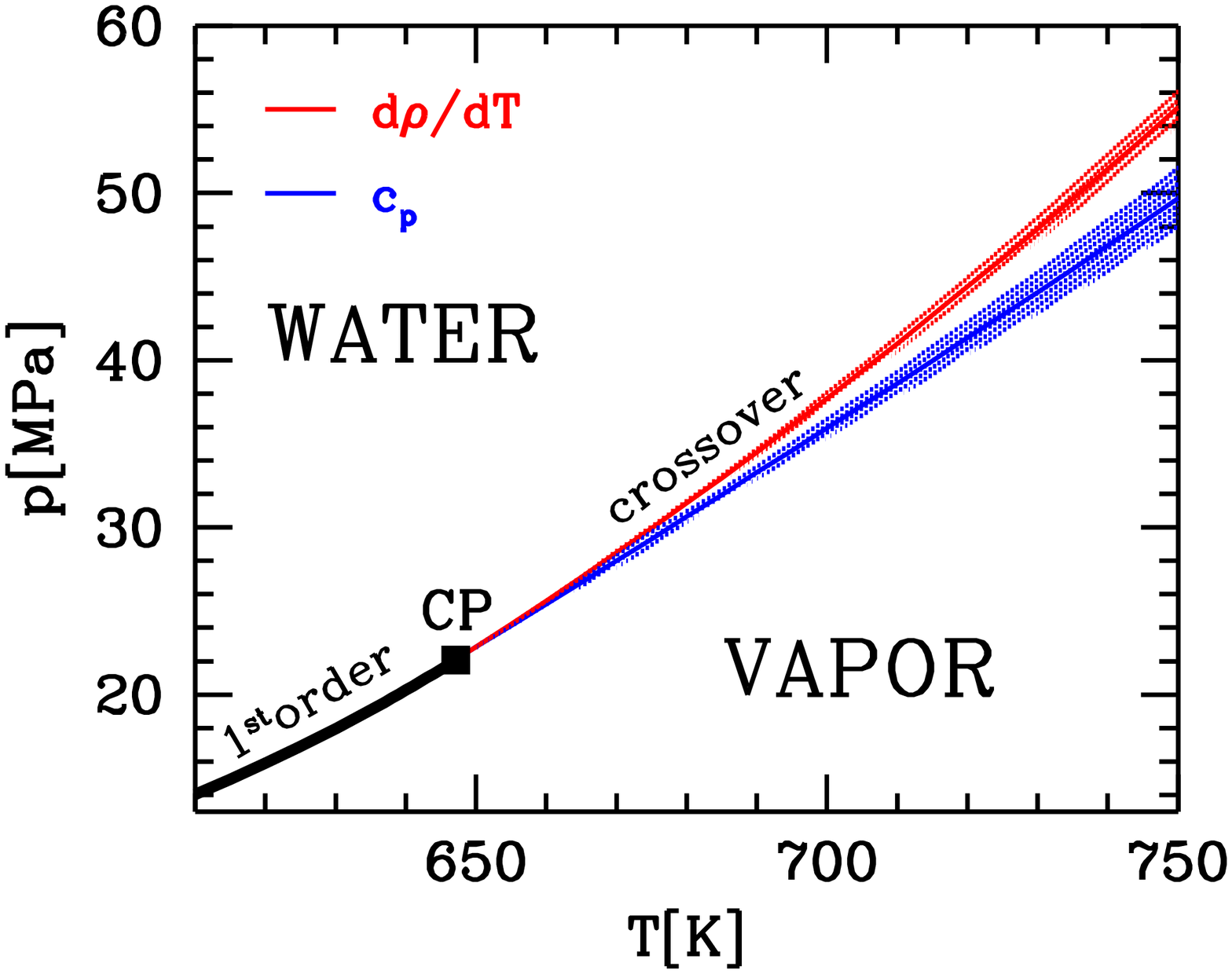}
\caption{\label{steam}
The water-vapor phase diagram. 
}
}

We carried out a finite size scaling analyses with the continuum
extrapolated height of the renormalized susceptibility. 
The renormalization of the chiral susceptibility
can be done by taking the second derivative of
the free energy density ($f$) with respect to the renormalized mass
($m_r$).
We apply the usual definition:
$f/T^4$ =$-N_t^4$\\
$\cdot [\log Z(N_s,N_t)/(N_t N_s^3)-\log Z(N_{s0},N_{t0})/(N_{t0} N_{s0}^3
)]$.
This quantity has a correct
continuum limit. The subtraction term is obtained at $T$=0, for which
simulations are carried out on lattices with $N_{s0}$, $N_{t0}$ spatial
and temporal extensions (otherwise at the same parameters of the action).
The bare light quark mass ($m_{ud}$) is related to $m_r$ by the
mass renormalization constant $m_r$=$Z_m$$\cdot$$m_{ud}$.
Note that $Z_m$
falls out of the combination
$m_r^2$$\partial^2$/$\partial$$m_r^2$=$m_{ud}^2$$\partial^2$/$\partial$$m_{ud}^2
$.
Thus, $m_{ud}^2\left[\chi(N_s,N_t)-\chi(N_{s0},N_{t0})\right]$ also has a
continuum limit
(for its maximum values for different $N_t$, and in the continuum limit
we use the shorthand
notation $m^2$$\Delta \chi$).

\begin{figure}\begin{center}
\centerline{\includegraphics[height=18cm, bb=190 160 410 710]{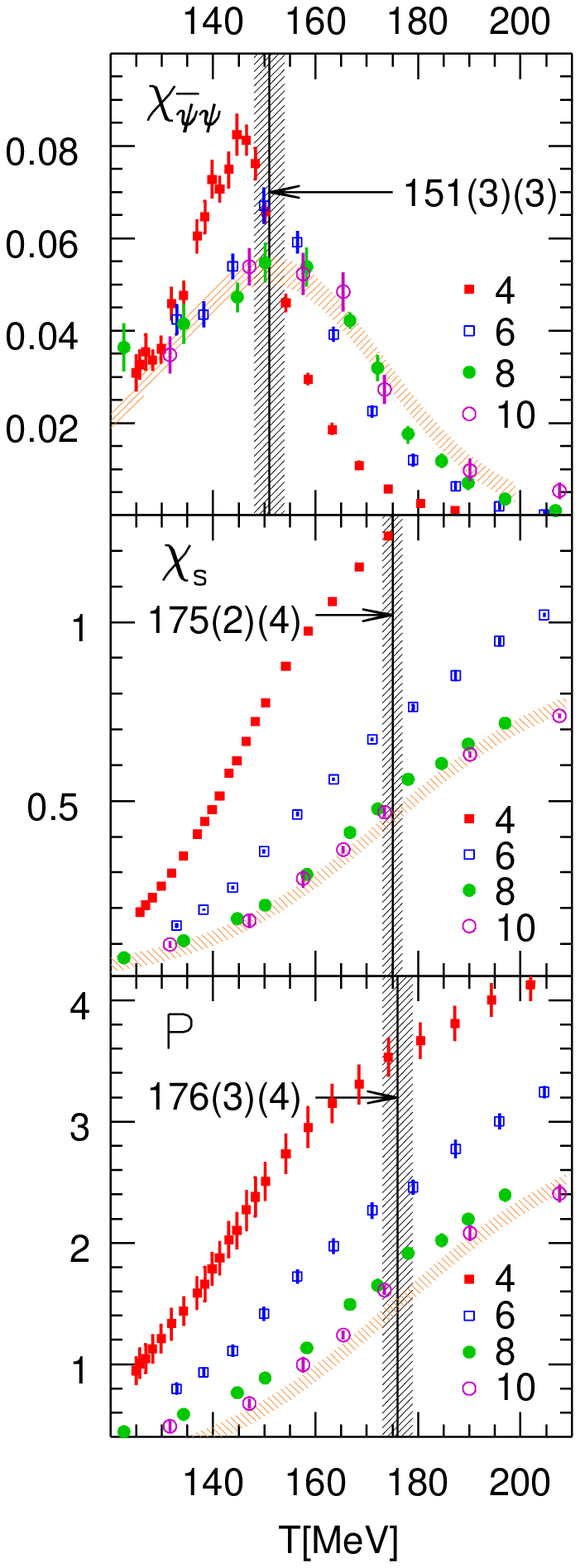}}
\end{center}\caption{\label{susc}
Temperature dependence of the renormalized
chiral susceptibility ($m^2\Delta \chi_{\bar{\psi}\psi}/T^4$), the strange
quark number susceptibility ($\chi_s/T^2$)
and the renormalized Polyakov-loop ($P_R$) in the transition region. The different symbols show the results for $N_t=4,6,8$ and
$10$ lattice spacings (filled and empty boxes for $N_t=4$ and $6$, filled and open circles for $N_t=8$
and $10$).
The vertical bands indicate the corresponding transition temperatures and
their uncertainties coming from the T$\neq$0 analyses. This error is
given by the number in the first parenthesis, whereas the error of the
overall scale determination is indicated by the number in the second
parenthesis. The orange bands show our continuum limit estimates for the
three renormalized quantities as
a function of the temperature with their uncertainties.
}
\end{figure}

In order to carry out the finite volume scaling in the continuum limit
we took three different physical volumes (see Figure \ref{cont_scal}).
The inverses of the volumes are shown in units of $T_c$.
For these 3 physical volumes we calculated the dimensionless combination
$T^4/m^2 \Delta \chi$ at 4 different
lattice spacings: 0.3fm was always off, otherwise the continuum
extrapolations could be carried out, which are shown on Figure \ref{cont_scal}. 
Our result is consistent with an approximately constant
behaviour, despite the fact that we had a factor of 5 difference in the
volume. The chance probabilities, that statistical fluctuations
changed the dominant behaviour of the volume dependence 
are negligible. As a conclusion we can say
that the staggered QCD transition at mu=0 is a cross-over.

\section{The transition temperature}

An analytic cross-over, like the QCD transition has no unique $T_c$. A particularly nice 
example for that is the water-vapor transition (c.f. Figure \ref{steam}). 
Up to about 650~K the transition 
is a first order one, which ends at a second order
critical point.  For a first or second order phase transition the 
different observables (such as density or heat capacity)
have their singularity (a jump or an infinitly high peak) at the same
pressure.  However, at even higher temperatures the transition is an
analytic cross-over, for which the most singular points are different. The
blue curve shows the peak of the heat capacity 
and the red one the inflection point of
the density. Clearly, these transition temperatures are different, which
is a characteristic feature of an analytic transition (cross-over). 

\FIGURE[t]{
\includegraphics[width=8cm,angle=0,bb=19 182 570 500]{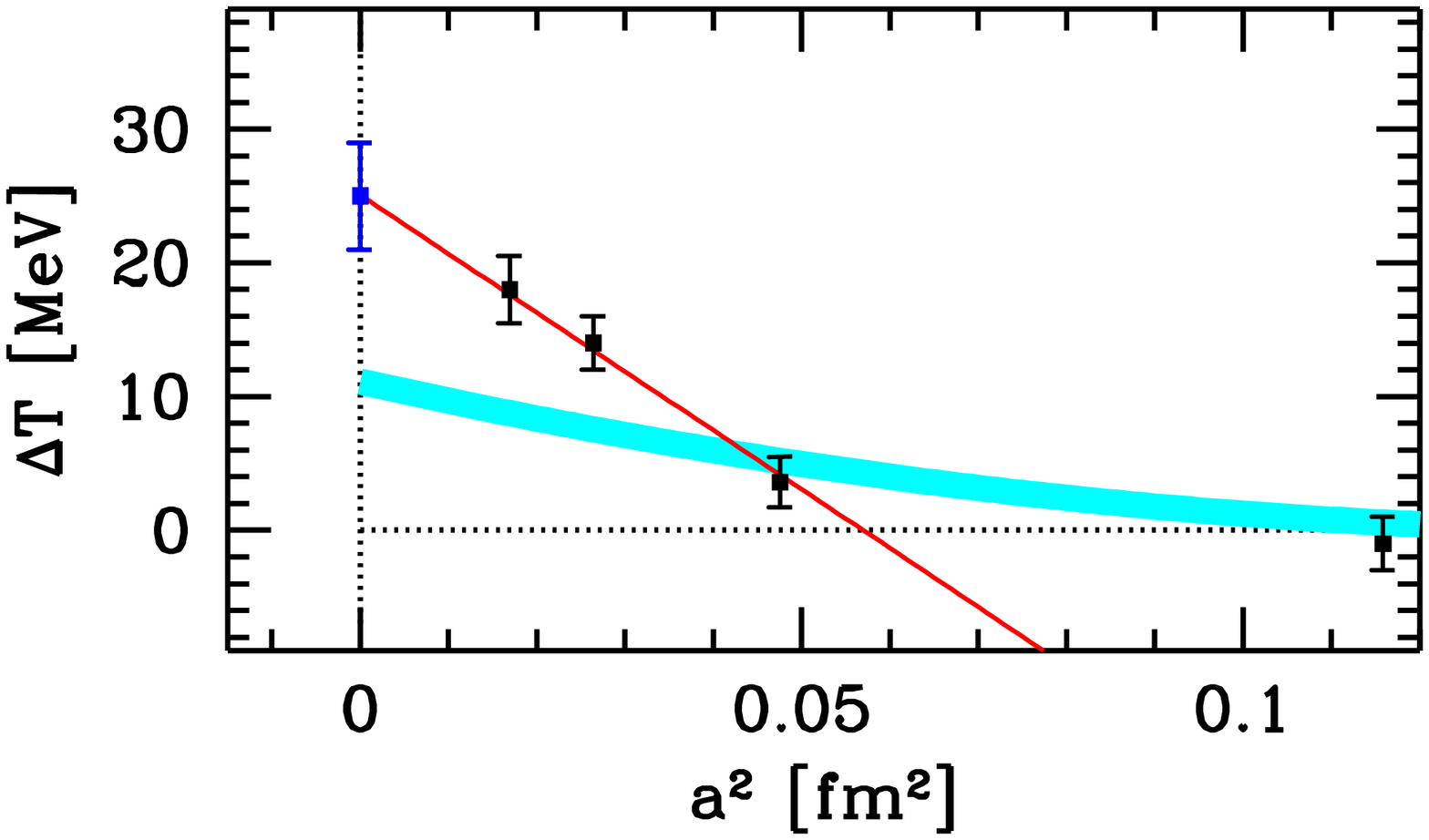}
\caption{\label{deltat}
Difference between the the $T_c$ values obtained
by the Polyakov loop and by the chiral condensate as a function of $a^2$.
}
}

In QCD we will study the
chiral and the quark number susceptibilities and the Polyakov loop. Usually
they give different $T_c$ values, but there is nothing wrong with it. 
As it was illustrated by the water-vapor transition it is a
physical ambiguity, related to the analytic behaviour of the transition.
There is another, non-phy\-si\-cal, ambiguity. If we used different observables
(such as the string tension, $r_0$, the rho mass or the kaon decay constant),
particularly at large lattice spacings we obtain different overall scales.
They lead to different $T_c$ values. 
This ambiguity disappeares in the continuum limit.
According to our experiences, at finite lattice spacings, 
the best choice is the kaon decay constant $f_k$. It is known
experimentally (in contrast to the string tension or $r_0$), thus no 
intermediate calculation with unknown systematics is involved. Furthermore
it can be measured on the lattice quite precisely.

Figure \ref{susc} shows the results for the chiral susceptibility, 
for the quark number
susceptibility and for the Polyakov loop. Red, blue, green, and purple
indicate $N_t$=4,6,8 and 10 lattices. $N_t$=4 is always off, the rest scales
nicely. The shaded regions indicate the continuum estimates. 
There is a surprising several sigma effect. The remnant of the chiral
transition happens at a quite different temperature than that of the
deconfining transition. It is quite a robust statement, since the Polyakov
transition region is quite off the $\chi$-peak, and the $\chi$-peak is quite far
from the inflection point of the Polyakov loop. This quite large differece
is also related to the fact that the transition is fairly broad. The widths are
around 30-40~MeV.

Due to the broadness of the transition the normalization prescription 
changes $T_c$, too. It is
easy to imagine why, just multiply a Gaussian by $x^2$ and the peak is
shifted. That means using $\chi/T^2$ gives about 10~MeV higher $T_c$ than
our definition, for which a $T^4$ normalization was applied. (Note, that
for the unrenormalized $\chi$ a $T^2$ normalization is natural, whereas
for the renormalized $\chi$ the natural normalization is done by $T^4$.
This kind of naturalness manifests itself as possibly small errors
of the observable.) 

Figure \ref{deltat} shows the difference between the $T_c$ values obtained 
by the Polyakov loop and by the chiral condensate as a function of the 
lattice spacing squared. The blue band indicates the difference for
the chiral susceptibility peak position for the $T^2$ and $T^4$
normalization. Thus using the $T^2$ normalization no difference can
be seen for $N_t$=4 and 6, a slight difference is observed for $N_t$=8
and a reliable continuum extrapolation needs $N_t$=6,8 and 10.

Our result on $T_c$ and that of the MILC Collaboration 
($T_c$=169(12)(4)~MeV \cite{Bernard:2004je}) are consistent within the
(quite sizable) errorbars. 

\begin{figure}\begin{center}
\includegraphics[width=13cm,angle=0,bb=0 0 765 85]{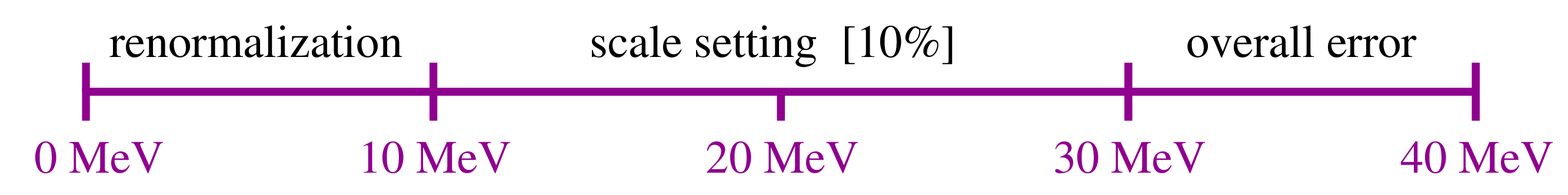}
\end{center}\caption{\label{tc_diff}
Possible contributions to the 40~MeV difference between the results
of Refs. \cite{Aoki:2006br} and \cite{Cheng:2006qk}.
}
\end{figure}

\begin{figure}\begin{center}
\includegraphics*[height=6.5cm,bb=300 165 592 420]{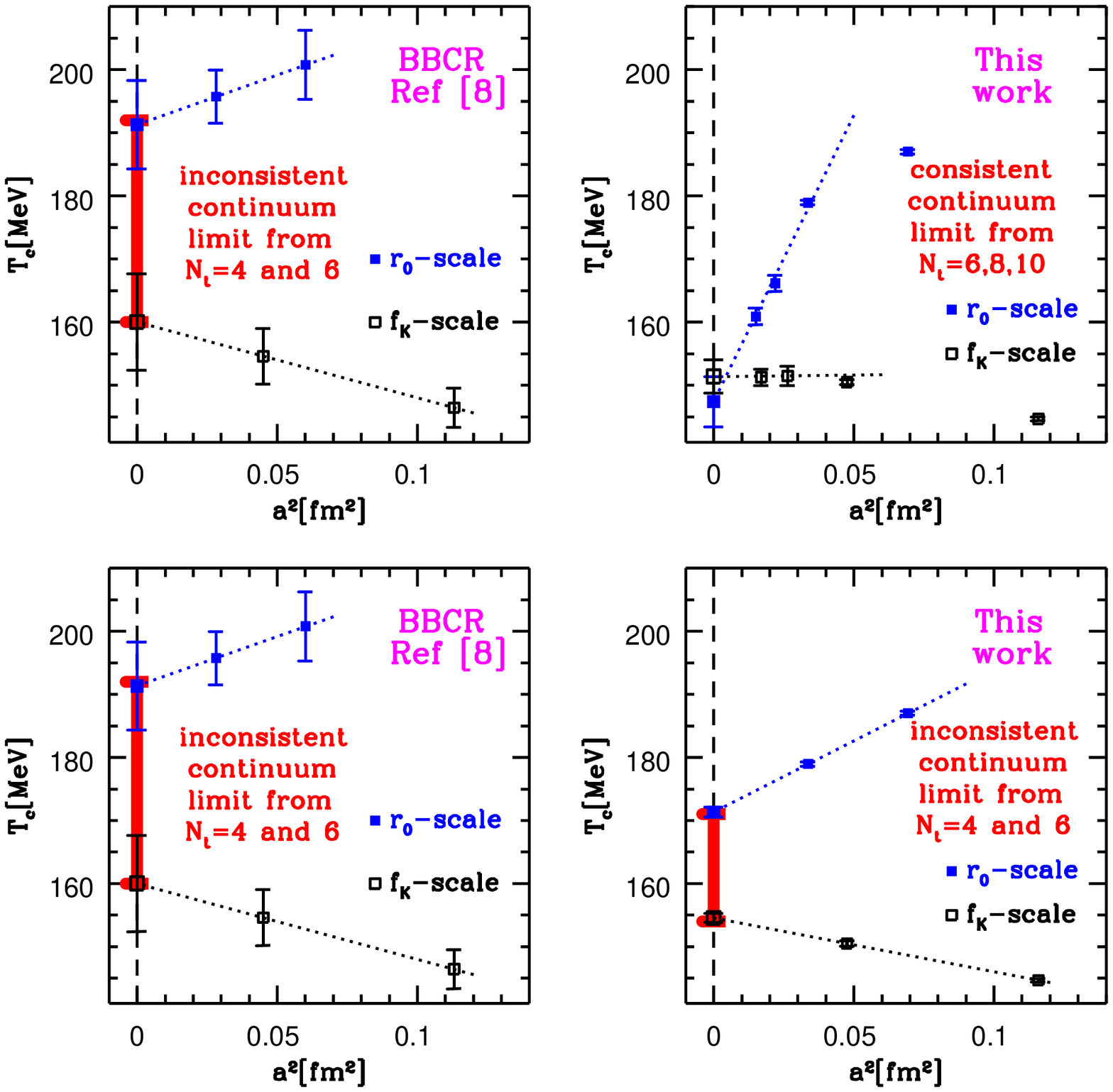}\hspace*{0.5cm}\includegraphics*[height=6.5cm,bb=300 440 592 695]{note3.eps}
\end{center}\caption{\label{note3}
Continuum extrapolations based on $N_t$=4 and 6 (left panel: inconsistent
continuum limit) and using  $N_t$=6,8 and 10 (right panel: consistent
continuum limit).  
}
\end{figure}

However, our result
contradicts the recent Bielefeld-Brookhaven-Columbia-Riken result 
\cite{Cheng:2006qk}, 
which obtained 192(7)(4)~MeV from both the
chiral susceptibility and Polyakov loop. This value 
is about 40~MeV larger than 
our result for the chiral susceptibility (for the Polyakov loop 
susceptibility the results agree within about 1.5$\sigma$). 
What are the differences between their analyses
and ours, and how do they contribute to the 40~MeV discrepancy? 
The most important contributions to the discrepancy are shown by 
Figure \ref{tc_diff}. The first difference is, that in \cite{Cheng:2006qk} no
renormalization was carried out, instead they used the unrenormalized quantity 
$\chi/T^2$. Due to the broadness of the distribution this observable
leads to about 10~MeV larger $T_c$ than our definition. The overall
errors can be responsible for another 10~MeV. The origin of the remaining
20~MeV is somewhat more complicated. One possible explanation can be
summarized as follows. 
In Ref. \cite{Cheng:2006qk} only $N_t$=4 and 6 were used, which correspond to
lattice spacings a=0.3 and 0.2~fm, or $a^{-1}$=700MeV and 1GeV. 
These lattices are
quite coarse and it seems to be obvious, that no unambiguous scale can be
determined for these lattice spacings. 
The overall scale in Ref. \cite{Cheng:2006qk} was set by $r_0$ and no
cross-check was done by any other quantity independent of the static
potential (e.g. $f_k$). This choice might lead to an ambiguity for the
transition temperature, which is illustrated for our data on Figure \ref{note3}.
Using only $N_t$=4 and 6 the continuum extrapolated 
transition temperatures are quite different
if one took $r_0$ or $f_K$ to determine the overall scale. This inconsistency
indicates, that these lattice spacing are not yet in the scaling region
(similar ambiguity is obtained by using the p4 action of \cite{Cheng:2006qk}).  
Having $N_t$=4,6,8 and 10 results 
this ambiguity disappears (as usual $N_t$=4 is off), 
these lattice spacings are already in the scaling region 
(at least within our accuracy). This phenomenon is not surprising at all.
As it was already mentioned e.g. the asqtad action 
at $N_t$$\approx$10 (which corresponds to about a=0.12~fm
lattice spacing) has $\approx$10\% scale difference 
predicted by $r_1$ or $f_K$. 

The ambiguity related to the inconsistent continuum limit is clearly
unphysical, and it is resolved as we take smaller and smaller lattice spacings
(c.f. Figure \ref{note3}). The differences between the $T_c$ values for 
different observables are physical, it is a consequence of the cross-over
nature of the QCD transition. There is another phenomenon, namely the volume
dependence of $T_c$, which is also physical. Recently A. Bazavov and B. Berg 
studied \cite{Bazavov:2007zz}
the volume dependence of $T_c$ in the pure SU(3) theory (for which
the transition is a first order phase transition). They applied the usual
periodic boundary condition (which approximates the thermodynamic limit
in a very effective way) and also a disorder wall boundary condition.
The difference between the $T_c$ values can be as large as 30~MeV (c.f.
Figure \ref{berg}). Clearly, a detailed study of this phenomenon 
is needed for dynamical QCD. 

\FIGURE[t]{
\includegraphics[width=7.5cm,angle=0]{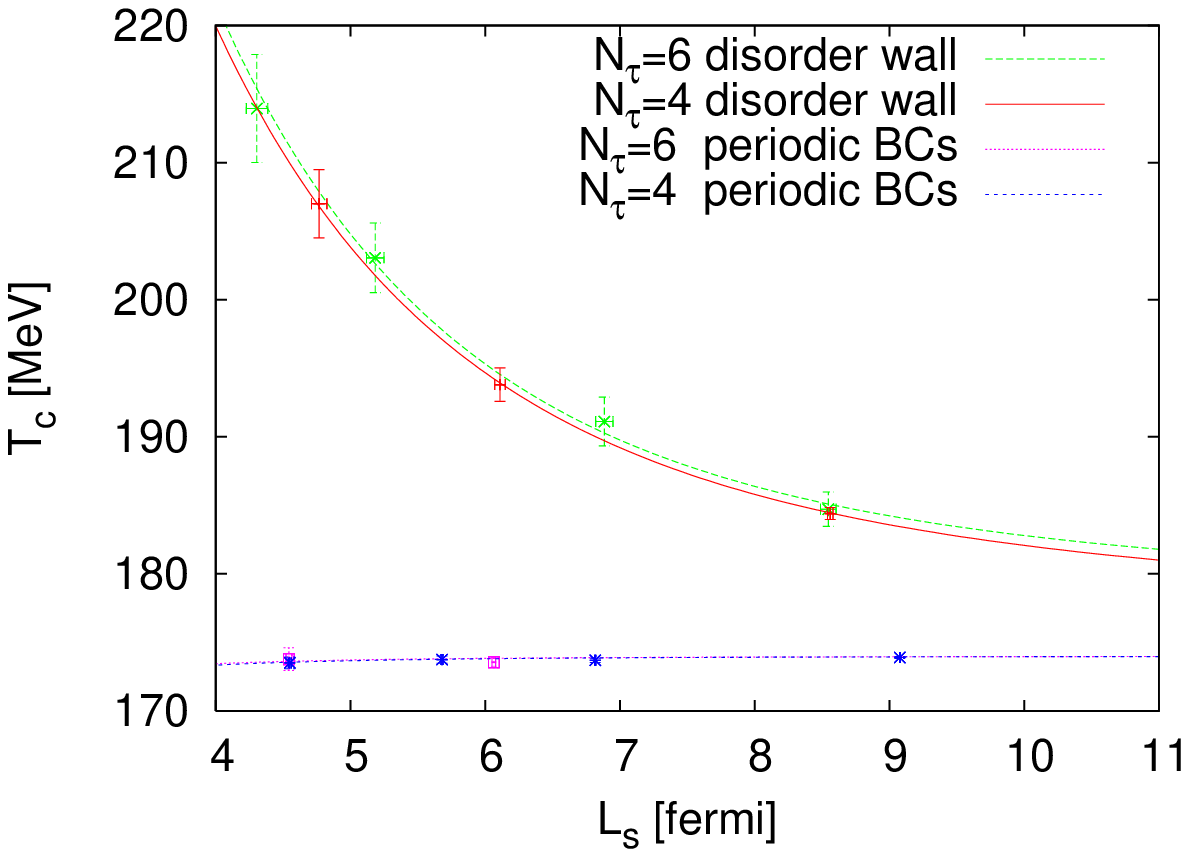}
\caption{\label{berg}
Transition temperature for the pure SU(3) theory as a function of the
spatial size for $N_t$=4 and 6. The lower, almost horizontal lines
represent the results with periodic boundary condition, whereas the  
upper curves show the results with disorder wall boundary condition
(Figure from Ref. \cite{Bazavov:2007zz}).}
}

\section{The renormalized static potential in the continuum limit}

Our aim is to compute the free energy of a static quark-antiquark pair. There
are several measurements on this quantity in the literature (for recent
publications cf.
\cite{Kaczmarek:2002mc,Bornyakov:2003ev,Kaczmarek:2004gv,Maezawa:2007fc}).
Here we go beyond these computations (for details see \cite{Fodor:2007mi}), 
we use physical quark masses and perform
a careful continuum limit extrapolation with the necessary renormalization
procedure.

The quark-antiquark free energy can be expressed
as correlators of Polyakov loops:
\begin{equation}
  e^{-F_{\bar{q}q}(\r)/T} \sim \sum\limits_\x \exv{\Tr P(\x)\,\Tr P^\dagger(\x+\r)},
\end{equation}
where \r\ is a vector in the spatial direction, $T=1/(N_t a)$ is the
temperature and $\x$ runs over all the spatial lattice sites. $P$ is the
Polyakov loop.

In full dynamical QCD at
large distances it is favorable to generate a quark-antiquark pair from the
vacuum, which then screens the color field between the two Polyakov
loops~\cite{Bali:2005fu}.  From this point (the string breaking scale) the
lowest energy level will be insensitive of the position of the heavy quarks,
resulting in a constant free energy.  

At finite temperature the above picture persists, but we can also have general
expectations about the temperature dependence. Physically we expect that in a
thermal vacuum it is easier to generate a quark-antiquark pair than at $T=0$,
since there
are thermally excited particles around which can scatter on the gluonic
string
between the static quark-antiquark pair. The gluonic string
can more easily break into a dynamical quark-antiquark pair.

When we approach the continuum limit, the value of the un-renormalized free
energy diverges. This is because in a single Polyakov loop the self-energy is
divergent. We expect:
\begin{equation}
  \exv{\Tr P(x)}\biggr|_\mathrm{div} = e^{-C(a) N_ta} = e^{-C(a)/T},
\end{equation}
where $C(a)\to\infty$ in the continuum limit. At finite $a$ the specific value
of $C(a)$ has no physical meaning, since it depends on how we define the
``divergent part'' of the self-energy (renormalization scheme).
In the literature there are several
ways to fix this constant \cite{Kaczmarek:2004gv,Aoki:2006br}.
A possible way of fixing $C(a)$ is to take a physical observable based on
$F_{\bar{q}q}$, and require that it should be independent of $a$.
In fact, the most useful quantity in our
calculation was the constant value of the free energy after the string
breaking/screening, at a fixed temperature (its value was set to 
$T_0=190$ MeV in 
the calculation). The value of $T_0=190$ MeV was
motivated by the fact that it is already in the deconfined phase
where the statistical errors of the free energy are much smaller than
in the confined phase. 
For the calculation we used the gauge configurations of Ref~\cite{Aoki:2006br}.
Having determined the renormalized free energy for all lattice spacings, we
could take the continuum limit by using the $N_t=4,6,8$ and 10 free energies,
and extrapolating in $1/N_t^2 \sim a^2 \to 0$. 
In the studied region $N_t=8$ and 10 results almost completely coincide.
Therefore a safe extrapolation to $1/N_t^2=0$ is possible. We estimate the
systematic error of this extrapolation by comparing the results coming from
$N_t=6,8,10$ extrapolation and $N_t=8,10$ extrapolation. The result for the
renormalized free energy at different temperatures, including both the
statistical and the systematic errors, can be seen on Figure \ref{pot}.

\FIGURE[t]{
\includegraphics[height=6.3cm,bb=15 180 570 610]{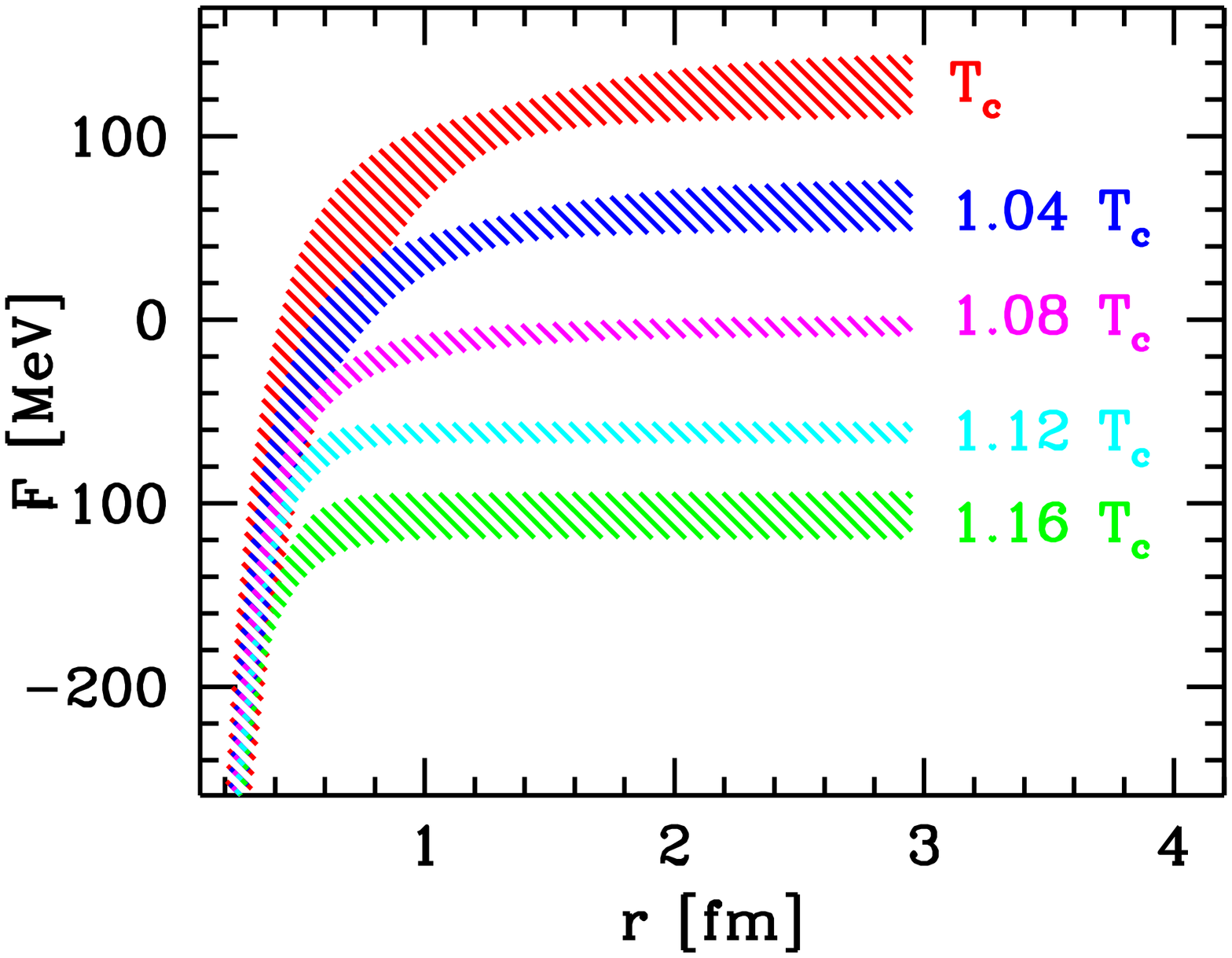}
\caption{\label{pot}
The renormalized free energies in the continuum limit.}
}

\section{The equation of state at high temperatures}

While available lattice results for the equation of state 
(both for pure gauge theory
 and for full QCD) end at around $5 \cdot T_c$, standard perturbation 
theory converges only at extremely high temperatures (at $5 \cdot T_c$,
the different perturbative orders can not even tell the sign of the 
deviation from the Stefan-Boltzmann li\-mit). Until recently no link 
between the two most systematic methods of QCD, namely perturbation
theory and lattice\\ QCD, existed for bulk thermodynamical quantities.

It is easy to understand
the difficulties for lattice calculations beyond these (e.g. $5 \cdot T_c$)
temperatures. The correctly renormalized pressure is obtained by
subtracting the vacuum term, thus 
$p_{\rm ren}(T)$=$p(T)$-$p(T=0)$,
where $p$=$T/V\cdot \log(Z)$ and $Z$ is the partition function. 
Since the partition function itself is difficult to determine, usually
the integral method is used, which e.g. for the plaquette action is given by
\begin{equation}
p_{\rm ren}(T)=p(T) - p(T=0)=\int d\beta \left( \langle {\rm Pl} 
\rangle_{T}-\langle {\rm Pl} \rangle_{0} \right)
\label{eq:p_int}
\end{equation}

As an example imagine a calculation at $20 \cdot T_c$ on $N_t$=8 lattices.
A back of an envelope estimate can convince you that it corresponds
to a lattice spacings of about 0.0075~fm or lattice sizes of
about 1000 in full QCD. This is clearly out of reach in the near future. 

In the perturbative approach the infrared properties of the non-abelian gauge
theories at non-vanishing temperatures are responsible for the bad 
convergence. 

\FIGURE[t]{
\includegraphics[height=6cm]{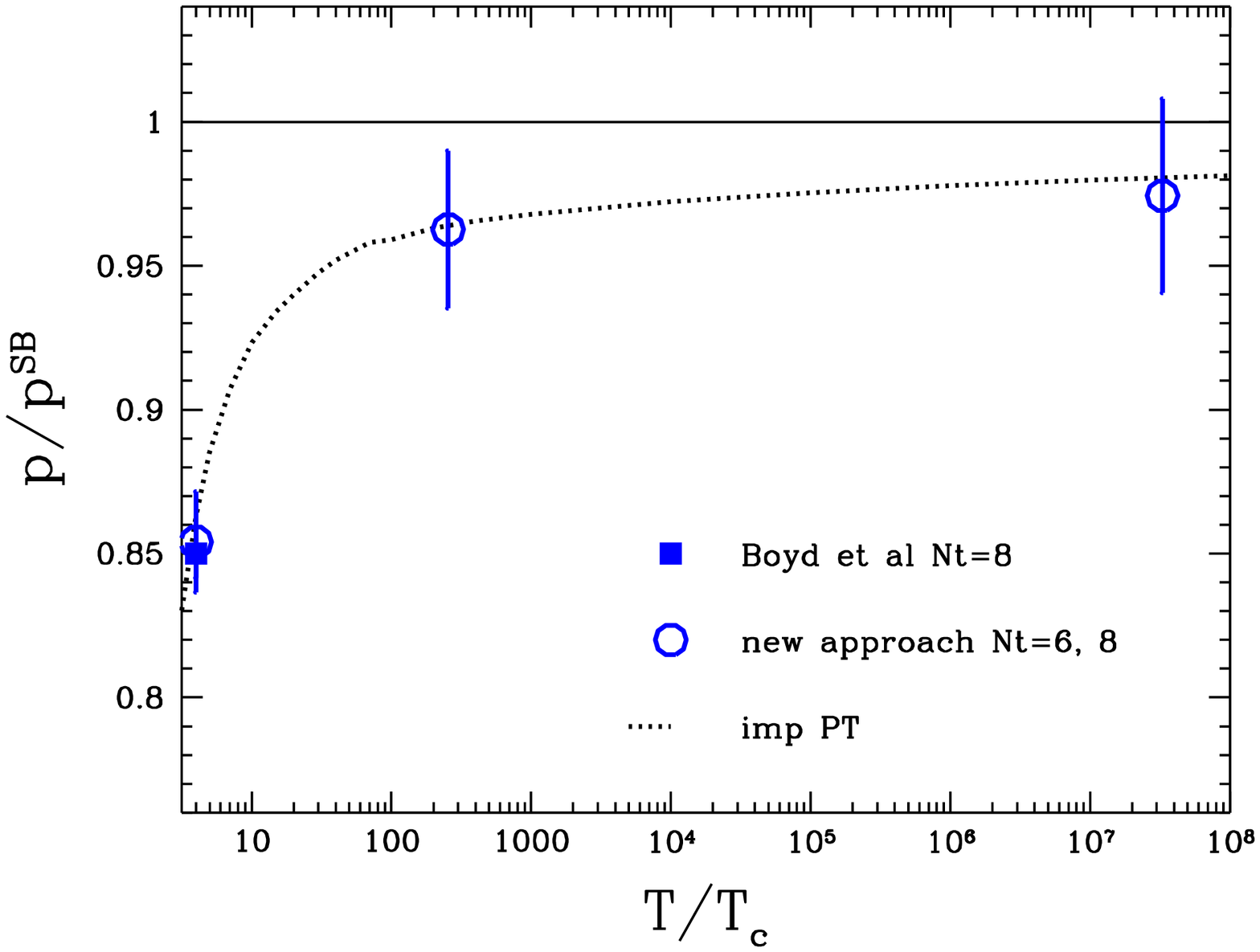}
\caption{\label{fig:p_new}
The pressure, normalized to its Stefan-Boltzmann value, as a 
function of the temperature obtained by our new technique. Results with 
small discretization errors ($N_t=8$, blue circles), seem to fit to the
improved perturbation theory prediction, 
and also reproduce results obtained by the 
standard method at lower temperatures. At the highest temperature, 
$3 \cdot 10^7 \cdot T_c$, the pressure (within its statistical uncertainty) 
is consistent with the Stefan-Boltzmann limit.}
}

To create a link between these two systematic approaches, 
we suggested a method \cite{Endrodi:2007tq}
to obtain lattice results for the pressure 
at temperatures, which were previously unreachable. It becomes possible 
to compare the data with perturbation theory formulae. The results are 
obtained using two different techniques: the first one is a new way to 
renormalize the pressure, the second one is a direct method 
to measure the pressure.
For illustration, the techniques and the results for pure SU(3) gauge theory
are presented. The extension to full QCD is straightforward.

The basic idea of the first technique is the observation, that the
the divergences that are removed by renormalization
are independent of the temperature. Thus the
subtraction of the vacuum term, which is needed for renormalization, can
be done at non-vanishing temperatures (actually even at very high
temperatures).
Thus, one can build up $p_{\rm ren}$ as a sum of differences
$p_{\rm ren}(T)$=[$p(T)$-$p(T/2)$]+[$p(T/2)$-$p(T/4)$]+$\dots$
In fact, we usually measure the dimensionless pressure, which can be 
obtained by including the $T^4$ factors. This leads to increasing
powers of 1/$2^4$ in the successive terms, thus only a few of them are 
needed. Applying this scheme, we may reach arbitrarily high temperatures 
using lattices with only $N_t$ and $2 \cdot N_t$ temporal extents. 
It is worth mentioning, that a similar formula can be constructed for 
the case of the normalized interaction measure 
$I \equiv (\epsilon - 3 \cdot p) / T^4$. 

Note, that independently of the renormalization 
procedure, a problem emerges within the integral method framework. Since 
strictly speaking, the pressure is only exactly zero at $T=0$, in principle 
we would have to carry out the integration starting from zero temperature. 
Thus, there is an uncertainty of setting the lower point of the integral.
This problem is solved by our direct method (c.f. \cite{Endrodi:2007tq}).

Let us consider the first bracketed term in $p_{\rm ren}(T)$. 
\begin{equation}
[p(T)-p(T/2)]=\frac{1}{N_tN_s^3}\log Z(N_t) - \frac{1}{2N_tN_s^3}\log Z(2N_t)=
\frac{1}{2N_tN_s^3}\log \left( \frac{Z(N_t)^2}{Z(2N_t)}\right).
\end{equation}
This can be determined by using $S_{1b}$ (action with one boundary in the
temporal direction), $S_{2b}$ (action with two boundaries in the
temporal direction) and an interpolating partition function
$\bar Z(\alpha)$=$\int \mathcal{D}U$
$exp[-(\alpha\cdot S_{2b}+(1-\alpha) \cdot S_{1b})]$
Using $\bar Z (\alpha)$, one obtains
\begin{equation}
[p(T)-p(T/2)] \sim \log \left( \frac{Z(N_t)^2}{Z(2N_t)}\right)=\log \left( \frac{\bar{Z}(1)}{\bar{Z}(0)}\right)=
\int_0^1 d\alpha \frac{d \log \bar{Z}(\alpha)}{d\alpha}=
\int_0^1 d\alpha \langle S_{1b} - S_{2b}\rangle
\label{eq:alpha}
\end{equation}
So we can calculate the pressure itself at any given temperature, without 
carrying out simulations at lower temperatures, and then performing an 
integral. (Note, however, that huge cancellations appear also within this framework). 

The method discussed above gives us the possibility to measure the pressure 
at very high temperatures. This was carried out using lattices with 
temporal extension $N_t=4$, $N_t=6$ and $N_t=8$. The temperature interval 
ranged from $4 \cdot T_c$ upto $3 \cdot 10^7 \cdot T_c$. Our results are 
shown on figure \ref{fig:p_new}. A comparison can be done with the 
standard method, namely the result of~\cite{Boyd:1996bx} for smaller 
temperatures. Here we present the results which are closest to the 
continuum limit, thus $N_t=8$. These results nicely follow the perturbative 
predictions ( for a recent paper see \cite{Laine:2006cp} and references 
therein). 

\section{Conclusion}

QCD thermodynamics results from the Budapest-Wuppertal group were summarized.
The necessary balance between $T=0$ and $T>0$ simulations was discussed in 
detail. As a consequence,
Symanzik improved gauge and stout-link improved staggered fermionic lattice
action was used in the simulations with an exact simulation algorithm.
Physical masses were taken both for the light quarks
and for the strange quark. The parameters were
tuned with a quite high precision,
thus at all lattice spacings the $m_K/f_k$ and $m_K/m_\pi$ ratios
were set to their experimental values with an accuracy better than 2\%.
Four sets of lattice spacings on lattices with 
$N_t$=4,6,8 and 10 temporal extensions were used 
(they correspond to lattice spacing 
$a$$\sim$0.3, 0.2, 0.15 and 0.12~fm) to carry out the continuum 
extrapolation. It
turned out that only $N_t$=6,8 and 10 can be used for a controlled
extrapolation, $N_t$=4 is out of the scaling region.

The nature of the T$>$0 transition was determined. The renormalized chiral
susceptibility was extrapolated to vanishing lattice
spacing for three physical volumes, the smallest and largest of
which differ by a factor of five. This ensures that a true transition
should result in a dramatic increase of the susceptibilities. No such
behaviour is observed: the finite-size scaling analysis showed that
the finite-temperature QCD transition in the hot early Universe 
was not a real phase transition, but an analytic crossover (involving
a rapid change, as opposed to a jump, as the temperature
varied). As such, it will be difficult to find experimental evidence
of this transition from astronomical observations. Since for present day
heavy ion experiments the baryonic chemical potential is also very small,
the above results apply for them, too.

The absolute scale for the T$>$0 transition was calculated.
Since the QCD transition is a non-singular cross-over
there is no unique $T_c$. This well-known phenomenon was 
illustrated on the water-vapor phase diagram.
Different observables led to different numerical $T_c$ values in the
continuum and thermodynamic limit also in QCD. Three observables were used 
to determine the corresponding transition temperatures. The peak of the
renormalized chiral
susceptibility predicted $T_c$=151(3)(3)~MeV, whereas $T_c$ based on the
strange quark number susceptibility resulted in 24(4)~MeV larger value.
Another quantity, which is related to the deconfining phase transition
in the large quark mass limit is the Polyakov loop. Its behavior predicted
a 25(4)~MeV larger transition temperature, than that of the chiral
susceptibility.
Another consequence of
the cross-over are the non-vanishing widths of the peaks even in the
thermodynamic limit, which were also determined. For the chiral susceptibility,
strange quark number susceptibility and Polyakov-loop we obtained widths of
28(5)(1)~MeV, 42(4)(1)~MeV and 38(5)(1)~MeV, respectively. 

The temperature dependent static potential was given. The same action and 
the same set of configurations were used as for the determination
of $T_c$. Since results
for different lattice spacings were available a careful renormalization
program was carried out.  

These features, numbers and functions are attempted
to be the full result for the $T$$\neq$0 transition, though other lattice
fermion formulations ---e.g.  Wilson fermions (for ongoing projects see e.g.
\cite{Maezawa:2007ew,Weinberg:2007tg}) or chiral fermions (for an early
dynamical overlap test see \cite{Fodor:2003bh}, for the domain wall approach
a recent presentation can be found in Ref.\cite{Vranas:2007})--- are needed 
to cross-check the findings with staggered fermions. 

A new technique is presented, which --in contrast to earlier
methods--  enables one to determine the equation of state at
very large temperatures. 
The method is based on the observation, 
that the divergences, which are removed by renormalization,
are independent of the temperature. Thus the
subtraction of the vacuum term, which is needed for renormalization, can
be done at non-vanishing temperatures (actually even at very high
temperatures).  A direct method was also suggested, which does not need 
an integration over the temperature.
Results for the pure SU(3) theory are presented
upto $3 \cdot 10^7 \cdot T_c$ temperatures. 

\section*{Acknowledgment}

Partial support of grants of \hbox{DFG F0 502/1}, \hbox{EU I3HP}, \hbox{OTKA
  AT049652} and \hbox{OTKA K68108} is acknowledged. The author thanks 
R. Hoffmann, S.D. Katz and K.K. Szabo for careful reading of the 
manuscript.


\begin{thebibliography}{99}
\bibitem{Karsch:2003va}
  F.~Karsch et al.,
  ``Where is the chiral critical point in 3-flavor QCD?,''
  Nucl.\ Phys.\ Proc.\ Suppl.\  {\bf 129} (2004) 614
  [arXiv:hep-lat/0309116].

\bibitem{Philipsen:2007}
O. Philipsen and P. De Forcrand, 
The chiral critical point of $N_f=3$ QCD: towards the continuum,
PoS(LATTICE 2007)178 

\bibitem{Endrodi:2007}
G. Endrodi, Z. Fodor, S.D. Katz and K.K. Szabo,  
The nature of the finite temperature QCD transition as a function 
of the quark masses,
PoS(LATTICE 2007)182 

\bibitem{Karsch:2007}
F. Karsch, Recent lattice results on finite temperature and density QCD, 
PoS(LATTICE 2007)015

\bibitem{Cheng:2006qk}
  M.~Cheng {\it et al.},
  ``The transition temperature in QCD,''
  Phys.\ Rev.\  D {\bf 74} (2006) 054507
  [arXiv:hep-lat/0608013].

\bibitem{Egri:2006zm}
  G.~I.~Egri, Z.~Fodor, C.~Hoelbling, S.~D.~Katz, D.~Nogradi and K.~K.~Szabo,
  ``Lattice QCD as a video game,''
  Comput.\ Phys.\ Commun.\  {\bf 177} (2007) 631
  [arXiv:hep-lat/0611022].

\bibitem{Aoki:2005vt}
  Y.~Aoki, Z.~Fodor, S.~D.~Katz and K.~K.~Szabo,
  ``The equation of state in lattice QCD: With physical quark masses  towards
  the continuum limit,''
  JHEP {\bf 0601} (2006) 089
  [arXiv:hep-lat/0510084].

\bibitem{Aoki:2006we}
  Y.~Aoki, G.~Endrodi, Z.~Fodor, S.~D.~Katz and K.~K.~Szabo,
  ``The order of the quantum chromodynamics transition predicted by the
  standard model of particle physics,''
  Nature {\bf 443} (2006) 675
  [arXiv:hep-lat/0611014].

\bibitem{Aoki:2006br}
  Y.~Aoki, Z.~Fodor, S.~D.~Katz and K.~K.~Szabo,
  ``The QCD transition temperature: Results with physical masses in the
  continuum limit,''
  Phys.\ Lett.\  B {\bf 643} (2006) 46
  [arXiv:hep-lat/0609068].

\bibitem{Bernard:2004je}
  C.~Bernard {\it et al.}  [MILC Collaboration],
  ``QCD thermodynamics with three flavors of improved staggered quarks,''
  Phys.\ Rev.\  D {\bf 71} (2005) 034504
  [arXiv:hep-lat/0405029].

\bibitem{Bazavov:2007zz}
  A.~Bazavov and B.~A.~Berg,
  ``Deconfining Phase Transition on Lattices with Boundaries at Low
  Temperature,''
  Phys.\ Rev.\  D {\bf 76} (2007) 014502
  [arXiv:hep-lat/0701007].

\bibitem{Kaczmarek:2002mc}
  O.~Kaczmarek, F.~Karsch, P.~Petreczky and F.~Zantow,
  ``Heavy quark anti-quark free energy and the renormalized Polyakov loop,''
  Phys.\ Lett.\  B {\bf 543} (2002) 41
  [arXiv:hep-lat/0207002].

\bibitem{Bornyakov:2003ev}
  V.~Bornyakov {\it et al.},
  ``Heavy quark potential in lattice QCD at finite temperature,''
  arXiv:hep-lat/0301002.

\bibitem{Kaczmarek:2004gv}
  O.~Kaczmarek, F.~Karsch, F.~Zantow and P.~Petreczky,
  ``Static quark anti-quark free energy and the running coupling at finite
  temperature,''
  Phys.\ Rev.\  D {\bf 70} (2004) 074505
  [Erratum-ibid.\  D {\bf 72} (2005) 059903]
  [arXiv:hep-lat/0406036].

\bibitem{Maezawa:2007fc}
  Y.~Maezawa, N.~Ukita, S.~Aoki, S.~Ejiri, T.~Hatsuda, N.~Ishii and K.~Kanaya
                  [WHOT-QCD Collaboration],
  ``Heavy-Quark Free Energy, Debye Mass, and Spatial String Tension at   Finite
  Temperature in Two Flavor Lattice QCD with Wilson Quark Action,''
  Phys.\ Rev.\  D {\bf 75} (2007) 074501
  [arXiv:hep-lat/0702004].

\bibitem{Fodor:2007mi}
  Z.~Fodor, A.~Jakovac, S.~D.~Katz and K.~K.~Szabo,
  ``Static quark free energies at finite temperature,''
  arXiv:0710.4119 [hep-lat].

\bibitem{Bali:2005fu}
  G.~S.~Bali, H.~Neff, T.~Duessel, T.~Lippert and K.~Schilling  [SESAM
                  Collaboration],
  ``Observation of string breaking in QCD,''
  Phys.\ Rev.\  D {\bf 71} (2005) 114513
  [arXiv:hep-lat/0505012].

\bibitem{Endrodi:2007tq}
  G.~Endrodi, Z.~Fodor, S.~D.~Katz and K.~K.~Szabo,
  ``The equation of state at high temperatures from lattice QCD,''
  arXiv:0710.4197 [hep-lat].

\bibitem{Boyd:1996bx} 
G.~Boyd, et al., 
``Thermodynamics of SU(3) Lattice Gauge Theory,'' 
 Nucl. Phys. {\bf B469} (1996) 419-444 
[arXiv:hep-lat/9602007].

\bibitem{Laine:2006cp} 
M.~Laine and Y.~Schroder, 
``Quark mass thresholds in QCD thermodynamics,'' 
Phys.\ Rev.\ D {\bf 73} (2006) 085009 
[arXiv:hep-ph/0603048].

\bibitem{Maezawa:2007ew}
  Y.~Maezawa {\it et al.},
  ``Thermodynamics and heavy-quark free energies at finite temperature and
  density with two flavors of improved Wilson quarks,''
  arXiv:0710.0945 [hep-lat].

\bibitem{Weinberg:2007tg}
  V.~Weinberg, E.~M.~Ilgenfritz, K.~Koller, Y.~Koma, Y.~Nakamura, G.~Schierholz and T.~Streuer
                  [DIK Collaboration],
  ``The chiral transition on a $24^3$10 lattice with $N_f$=2 clover sea quarks
  studied by overlap valence quarks,''
  arXiv:0710.2565 [hep-lat].

\bibitem{Fodor:2003bh}
  Z.~Fodor, S.~D.~Katz and K.~K.~Szabo,
  ``Dynamical overlap fermions, results with hybrid Monte-Carlo algorithm,''
  JHEP {\bf 0408}, 003 (2004)
  [arXiv:hep-lat/0311010].

\bibitem{Vranas:2007}
P. Vranas,
``The $N_t$=8 QCD thermal transition with DWF,''
PoS(LATTICE 2007)235. 

\end{thebibliography}
\end{document}